\documentclass[aps,prl,preprint,groupedaddress]{revtex4}
\usepackage{graphicx}
\usepackage{amsmath,amsfonts,amssymb}
\usepackage{color}
\usepackage{float}
\begin{document}

\title {Ricci linear Weyl/Maxwell mutual sourcing}

\author{Aharon Davidson}
\email{davidson@bgu.ac.il}
\homepage{https://physics.bgu.ac.il/~davidson/}
\author{Tomer Ygael}
\email{tomeryg@post.bgu.ac.il}
\affiliation{Physics Department, Ben-Gurion
University of the Negev, Beer-Sheva 84105, Israel}

\begin{abstract}
We elevate the field theoretical similarities between
Maxwell and Weyl vector fields into a full local
scale/gauge invariant Weyl/Maxwell mutual sourcing
theory.
In its preliminary form, and exclusively in four dimensions,
the associated Lagrangian is dynamical scalar field free,
hosts no fermion matter fields, and Holdom kinetic
mixing is switched off.
The mutual sourcing term is then necessarily spacetime
curvature (not just metric) dependent, and inevitably
Ricci linear, suggesting that a non-vanishing spacetime
curvature can in principle induce an electromagnetic
current. In~its mature form, however, the~Weyl/Maxwell mutual
sourcing idea serendipitously constitutes a~novel variant
of the gravitational Weyl-Dirac (incorporating Brans-Dicke)
theory.
Counter intuitively, and again exclusively in four dimensions,
the optional quartic scalar potential gets consistently
replaced by a Higgs-like potential, such that the co-divergence
of the Maxwell vector field resembles a conformal vacuum
expectation value.

\textbf{Essay awarded Honorable Mention by the Gravity Research
Foundation 2020.}
\end{abstract}

\maketitle

Within the framework of Riemann geometry, with~tensor
fields serving as the fundamental objects, ordinary derivatives
are consistently replaced by covariant derivatives to assure
diffeomorphism invariance.
Going one step further into the territory of Weyl geometry
~\cite{Weyl1,Weyl2},
the tensor fields are traded for so-called co-tensor fields,
and the covariant derivatives are generalized into co-covariant
(also known as starred $\star$) derivatives,
respectively.
Resembling a $U(1)$ local gauge theory, the~star derivation
procedure mandatorily introduces a new player into the
game, the~Weyl vector field $a_\mu (x)$.
The~prototype theory in this category is Weyl-Dirac gravity
~\cite{Dirac,Utiyama,Cheng}, a~local scale symmetric generalization
of Brans-Dicke theory~\cite{BD}.
A counter example is provided by $C^2$ conformal gravity
~\cite{C2,tHooft} which, owing to the Weyl tensor
$C^\mu_{~\nu\lambda\sigma}$ being an in-tensor (co-tensor
of weight zero), solely in 4-dimensions, does~not require the
presence of $a_\mu (x)$.
Other theoretical directions include two scalar
gravity-anti-gravity theories~\cite{GaG1,GaG2,GaG3,GaG4}, and~Kaluza-Klein
reduced higher dimensional local scale symmetric theories
~\cite{conKK1,conKK2,conKK3}.
However, while treated on equal canonical footing in the
Lagrangian formalism, Maxwell vector field $A_\mu (x)$
and Weyl vector field $a_\mu (x)$ play completely different
roles in theoretical~physics.

From the geometric point of view, the~differences
between these two vector fields sharpens.
While~$A_\mu (x)$ constitutes an in-vector, $a_\mu (x)$
does not constitute a co-vector at all.
However, in~spite of their physical and geometrical differences,
these two vector fields do share a similar transformation law
under their corresponding local symmetries.
To be specific,
\begin{eqnarray}
& A_\mu (x) \rightarrow A_\mu (x)
-\partial_\mu \varPhi (x)~,&
\label{Alaw}\\
& a_\mu (x) \rightarrow a_\mu (x)
-\partial_\mu\varphi (x) ~,&
\label{aWlaw}
\end{eqnarray}
with $\varPhi$ taking values on a circle whereas $\varphi$
on the real line.
In turn, both their kinetic terms, namely $F_{\mu\nu}$
and $f_{\mu\nu}$ respectively,  transform alike as Weyl
in-scalars, and~a Holdom-style kinetic mixing~\cite{Holdom}
becomes then field theoretically permissible.
In this essay, however, with~or without invoking the kinetic
mixing term, we elevate the apparent field theoretical
similarities between Maxwell and Weyl vector fields into
a full local scale/gauge invariant mutual sourcing theory.
In line with Einstein-Hilbert and especially with Weyl-Dirac
actions, and~solely in a 4-dimensional spacetime, we show
that the scale symmetric Weyl/Maxwell mutual
source mixing is necessarily spacetime curvature dependent
(not just metric dependent), and~inevitably Ricci linear.
This way, a~non-vanishing spacetime curvature becomes an
unconventional source of the electromagnetic~current.

Let our starting point be the familiar 4-dimensional action
involving a linear electromagnetic coupling term
\begin{equation}
{\cal I}_{EM}
=\int \left({\cal L}_G-\frac{1}{4}F^2
-J^\mu A_\mu\right)
\sqrt{-g}~d^4 x ~,
\label{em}
\end{equation}
with $J^\mu$ serving as the external electromagnetic
source current, and~${\cal L}_G$ denoting the yet
unspecified gravitational part of the Lagrangian.
{To keep gauge invariance manifest already at the
Lagrangian level, one~may invoke a Lagrange multiplier
$\eta$, and~simply replace $A_\mu$ by
$A_\mu-\eta_{;\mu}$, such that $\eta\rightarrow \eta+\Phi$}.
It is only at the stage when gauge fixing becomes
permissible, e.g.,~at the level of the equations of motion,
that one may set $\eta=0$.
As dictated by the self consistency of the
associated Maxwell equations
$F^{\mu\nu}_{~;\nu}=J^\mu$, and~directly by the
variation with respect to $\eta$, the~theory maintains
gauge invariance only provided $J^\mu$ is locally
conserved $J^\mu_{~;\mu}=0$.
The action Equation~(\ref{em}) is furthermore local scale invariant
if ${\cal L}_G$ is such, and~if $J_\mu$ happens to be a
co-covariant vector of power
\begin{equation}
[J_\mu]=-2  ~~\Longleftrightarrow~~
[J^\mu]=-4~,
\end{equation}
where in our Weyl-Dirac notations,
\begin{eqnarray}
& [g_{\mu\nu}]=2 ~,~ [g^{\mu\nu}]=-2
~\Longrightarrow~ [\sqrt{-g}]=4 ~,& \\
& [A_{\mu}]=0 ~,~ [A^{\mu}]=-2
~\Longrightarrow~ [F_{\mu\nu}]=0 ~.&
\end{eqnarray}

The last formula deserves some attention.
Consider a co-covariant vector $V_\mu$ of power
$[V_\mu]=n$,
and recall that its covariant derivative $V_{\mu ;\nu}$
does not form a co-tensor.
Alternatively, one invokes a~co-covariant starred
derivative, and~show that the corresponding co-tensor
role is then taken by
\begin{equation}
V_{\mu \star \nu}=V_{\mu ;\nu}
-(n-1)a_\nu V_\mu+a_\mu V_\nu
-g_{\mu\nu} a^\lambda V_\lambda ~.
\label{V}
\end{equation}

In particular, notice the antisymmetric combination
\begin{equation}
V_{\mu \star \nu}-V_{\nu \star \mu}
=V_{\mu;  \nu}-V_{\nu; \mu}
+n(a_\mu V_\nu-a_\nu V_\mu)~,
\end{equation}
telling us that antisymmetric
$F_{\mu\nu}=A_{\mu;  \nu}-A_{\nu; \mu}
=A_{\mu \star \nu}-A_{\nu \star \mu}$ is in fact
an in-tensor simply because $A_\mu$ is an in-vector
(meaning $n=0$) to start with.
By the same token, if~$U^\mu$ is a co-contravariant
vector of power $[U^\mu]=n$, its star derivative is
given by
\begin{equation}
U^\mu_{\star \nu}=U^\mu_{;\nu}
-(n+1)a_\nu U^\mu+a^\mu U_\nu
-g^\mu_{~\nu} a^\lambda U_\lambda ~.
\label{V}
\end{equation}

In particular,  its co-divergence is given by
\begin{equation}
U^\mu_{\star\mu}=U^\mu_{;\mu}
-(n+4)a_\mu U^\mu ~.
\label{div}
\end{equation}

It is only for the special case of $n=-4$, that
we face the advantage of
$J^\mu_{\star\mu}=J^\mu_{;\mu}$.

The fact that $A_\mu$ and $a_\mu$ share
similar transformation laws under their
corresponding local symmetries, and~exhibit
kinetic terms of the one and the same structure,
may prematurely suggest, in~analogy with
Equation~(\ref{em}), an~action $\acute{\text a}$ la
\begin{equation}
\int \left[{\cal L}_G-\frac{1}{4}f^2
-j^\mu a_\mu \right]\sqrt{-g}~d^4 x ~.
\label{false}
\end{equation}

The trouble is that, while
$f_{\mu\nu}=a_{\mu;\nu}-a_{\nu;\mu}$ turns
out to be a legitimate in-tensor, the~Weyl vector
$a_{\mu}$ itself does not transform like a
co-vector at all.
Unlike the Weyl vector which transforms a la
Equation~(\ref{aWlaw}), a~power $n$ co-vector gets scaled
by a factor $e^{n\varphi (x)}$.
This is to say that the action Equation~(\ref{false}) is not
invariant under arbitrary local scale~transformations.

In search of a tenable coupling term to replace
the problematic $j^\mu a_\mu$, we first recall
that co-covariant starred derivatives are generically
linear in $a_{\mu}$.
For example, let $S$ be a co-scalar of power $n$,
then
\begin{equation}
S_{\star \mu}=S_{;\mu}-n a_\mu S~,
\end{equation}
with the bonus of having $[S_{\star \mu}]=n$ as well.
Thus, a~coupling term of the form
\begin{equation}
{\cal L}_{int}= j^\mu S_{\star \mu}
\propto  j^\mu a_\mu+...
\end{equation}
can certainly do, but~only provided
(i) $n\neq 0$ on self consistency grounds, and~(ii) The source current $j^\mu$ must constitutes
a co-vector of the exact power
\begin{equation}
[j^\mu]=-(n+4) ~~\Rightarrow
~~ [j_\mu]=-(n+2) ~.
\label{jn}
\end{equation}

Now, aiming towards Weyl/Maxwell mutual sourcing,
one would  like to identify $j_\mu$ with $A_\mu$.
This~is our goal, but~for this to be the case, recalling
that $[A_\mu]=0$,  we must first find a suitable
candidate for $S$, such that
\begin{equation}
[A_\mu]=0  ~\Longrightarrow~   [S]=-2 ~.
\label{S}
\end{equation}

What are the options?

At this stage, fundamental scalar fields are yet
to be introduced.
In fact, the~option of not introducing fundamental
scalar fields into the theory is exclusively viable in
four spacetime dimensions.
So, in~the absence of scalar fields, the~answer to
the above question must come from the geometry
of the underlying 4-dim curved spacetime.
The simplest curvature scalar to think of is no
doubt the Ricci scalar $R$.
However, unfortunately, $R$ cannot enter the game
as is, but~must be traded for its $\tilde{R}$
scale symmetric co-scalar variant.
In 4-dimensions, it is given by
\begin{equation}
\tilde{R} =R+6a^\mu_{;\mu}-6a^\mu a_\mu  ~.
\label{tildeR}
\end{equation}

Note that we prefer the notation $\tilde{R}$,
instead of the original $R^\star$ or $^\star R$,
leaving the star symbol solely for denoting
co-derivation.
The crucial observation now is that $[\tilde{R}]=-2$,
and the same is true for its~co-derivative
\begin{equation}
\tilde{R}_{\star\mu}=\tilde{R}_{;\mu}
+2a_\mu \tilde{R} ~.
\label{Rstarmu}
\end{equation}

In turn, the~master requirement Equation~(\ref{S}) can
now be  satisfied by naturally choosing $S=\tilde{R}$.
It is straightforward to verify that other powers of
$\tilde{R}$, as~well as higher order curvature co-scalars,
such~as $\tilde{R}^{\mu\nu}\tilde{R}_{\mu\nu}$ and
$\tilde{R}^{\mu\nu\lambda\sigma}
\tilde{R}_{\mu\nu\lambda\sigma}$, will not~do.

We can now close the circle.
Rather than assigning external non-dynamical source
currents $J^\mu$ and $j^\mu$, we let the Maxwell
vector field $A_\mu$ and the Weyl vector field $a_\mu$
source each other.
The result is the simplest dynamical scalar free local gauge/scale
invariant Weyl/Maxwell mixing theory described by the
action
\begin{equation}
I=-\int\left[\frac{1}{4}F^2+\frac{1}{4}f^2
+\frac{1}{2}e A^\mu \tilde{R}_{\star\mu}\right]
\sqrt{-g}~d^4x ~,
\label{Lmix}
\end{equation}
where $e$ is a universal dimensionless constant.
Recall that, in~analogy with the note following
Equation~(\ref{em}), $A_\mu$ is to be replaced
by $A_\mu-\eta_{\star\mu}=A_\mu-\eta_{;\mu}$,
with $[\eta]=0$,
whenever is needed (like here) to make gauge invariance
manifest already at the Lagrangian level.
It is crucial to notice that $A^\mu \tilde{R}$
happens to be a co-contravariant vector of the
special power $[A^\mu \tilde{R}]=-4$.
Hence, by~recalling Equation~(\ref{div}) twice, we find
\begin{equation}
A^\mu \tilde{R}_{\star\mu}
=-A^\mu_{\star\mu}\tilde{R}
+(A^\mu\tilde{R})_{\star\mu}
= -(A^\mu_{;\mu}-2A^\mu a_\mu)\tilde{R}
+(A^\mu\tilde{R})_{;\mu} ~.
\end{equation}

Up to a total divergence, and~by no coincidence,
also up to a total co-divergence, Equation~(\ref{Lmix}) can be
now re-written in the attractive $\tilde{R}$-linear form
\begin{equation}
{ I_0=-\int\left[\frac{1}{4}F^2+\frac{1}{4}f^2
-\frac{1}{2}e A^\mu_{\star\mu}\tilde{R} \right]
\sqrt{-g}~d^4x}
\label{Rlinear}
\end{equation}

The Weyl/Maxwell mutual sourcing can take a more
conventional form by introducing yet a~non-dynamical
real scalar field $\phi$ (accompanied by a suitable
Lagrange multiplier $\lambda$, such that $[\lambda]=-2$)
into the theory
\begin{equation}
I_1=-\int\left[\frac{1}{4}F^2
+\frac{1}{4}f^2-\phi^2\tilde{R}
+\lambda\left(\phi^2
-\frac{1}{2}e A^\mu_{\star\mu}\right)\right]
\sqrt{-g}~d^4x ~.
\end{equation}

However, for~the scalar field to become dynamical, a~supplementary in-scalar kinetic term is mandatory,
and following Dirac, the~Brans-Dicke coefficient
$\omega$ of such a term can be fully arbitrary, not
necessarily critical.
This leads us to
\begin{equation}
I_2=-\int\left[\frac{1}{4}F^2
+\frac{1}{4}f^2-\phi^2\tilde{R}
+\omega g^{\mu\nu}
\phi_{\star\mu}\phi_{\star\nu}
+\lambda\left(\phi^2
-\frac{1}{2}e A^\mu_{\star\mu}\right)\right]
\sqrt{-g}~d^4x ~,
\end{equation}
which, up~to the $\lambda$-term, establishes contact
with the Weyl-Dirac~theory.

With such an observation in hand, the~latest action
needs not be the final word, as~the Weyl-Dirac theory
is known to further allow for a quartic scalar potential.
Consequently, we cannot resist replacing the quadratic
$\phi^2$-constraint by a quartic $\phi^4$ potential, and~by consistently doing so, trading the auxiliary co-scalar
$\lambda$ for a dimensionless constant coefficient
$\Lambda$.
The resulting theory {reads} 
\begin{equation}
{I_3=-\int\left[\frac{1}{4}F^2
+\frac{1}{4}f^2-\phi^2\tilde{R}
+\omega g^{\mu\nu}
\phi_{\star\mu}\phi_{\star\nu}
+\Lambda \left(\phi^2
-\frac{1}{2}e A^\mu_{\star\mu}\right)^2\right]
\sqrt{-g}~d^4x}
\end{equation}
with $e=0$ signaling the exact Weyl-Dirac limit.
In fact, and~perhaps counter intuitively,
\begin{equation}
v^2 \equiv \frac{1}{2}e A^\mu_{\star\mu}=
e(\frac{1}{2}A^\mu_{;\mu} -a_\mu A^\mu)
\end{equation}
highly resembles (and can be referred to as) a
{conformal vacuum expectation value}.
The former constraint
$\phi^2=\frac{1}{2}e A^\mu_{\star\mu}$ is now
realized as the minimum (for $\Lambda>0$) of a
tenable Higgs potential.
We~note here again that, in~all evolving action versions
$I_{0,1,2,3}$, in~order to make gauge invariance
manifest already at the Lagrangian level, one
consistently replaces $A_\mu$ by
$A_\mu-\eta_{\star\mu}=A_\mu-\eta_{;\mu}$.
While~the presence of the $\eta$ is mandatory
as long as the $U(1)$ coupling is non-minimal, it can
eventually be integrated out by gauge fixing.
The situation may look somewhat reminiscent of
the Stueckelberg action for a massive vector field,
but recall that the present theory is a priori free
of any mass scale.
As~far as the physical interpretation of $\eta$ is
concerned, it should be clarified that it cannot be
regarded a~new independent dynamical scalar field.
The reason being that it is just the one and only
combination $A_{\mu}-\eta_{;\mu}$ which actually
enters the Lagrangian.

We modestly aimed towards Weyl/Maxwell mutual
sourcing, and~have automatically been driven into its
unified Weyl/Dirac/Maxwell/Higgs embedding.
Gravity just cannot stay out of the game.
There may be however a price to pay.
While the situation is apparently somewhat similar
to the Weyl-Dirac theory, the~differential equations
of motion are beyond second order (a counter
example is provided by the non-trivial local scale
invariant extension of the 4-dim Gauss-Bonnet theory).
If this is the case, then the Ricci linear coupling may
introduce ghosts and render the minimal theory sick.

It has not escaped our attention that, while sticking
to 4-dimensions, one is always free to add curvature
quadratics terms, for~example ${\cal L}_G=\tilde{R}^2$
or ${\cal L}_G=C^2$ without violating local scale
invariance.
Another pretentious attempt would be to add
Equation~(\ref{Rlinear}) to the standard
Einstein-Hilbert ${\cal L}_G=R$, which obviously
does not respect Weyl scale symmetry.
This would mean revising Einstein-Maxwell into
Einstein-Weyl/Maxwell theory,
and modifying even the Reissner-Nordstrom
solution accordingly.
Such generalizations are however beyond the scope
(and even beyond the rationale) of the present paper.
On pedagogical and simplicity grounds, however,
we hereby set ${\cal L}_G=0$ and first study the action
Equation~(\ref{Rlinear}) on its own~merits.

At any rate, here are some distinctive features of
the simplest Weyl/Maxwell mutual sourcing theory
$I_0$ prescribed by the action Equation~(\ref{Rlinear}):

\begin{itemize}
\item	The highlight is, roughly speaking, the~construction
of the Maxwell conserved current $J_{\mu}$ from
spacetime curvature (involving $a_\mu$ dependence).
The variation with respect to $A_\mu$ is straight
forward, giving rise to the conformal conservation law
\begin{equation}
\left(F^{\mu\nu}-
\frac{1}{2}e g^{\mu\nu}\tilde{R}\right)_{~\star\nu}=0 ~,
\end{equation}
where one can make use of the identity
$F^{\mu\nu}_{~\star\nu}=F^{\mu\nu}_{~;\nu}$.
Self consistency (and $g^{\mu\nu}_{~\star\nu}=0$)
then dictates the complementary co-scalar constraint
\begin{equation}
g^{\mu\nu}\tilde{R}_{\star\mu\star\nu}=0~.
\label{comp}
\end{equation}

Here again, owing to $[\tilde{R}_{\star \mu}]=-2$,
one can take advantage of
$g^{\mu\nu}\tilde{R}_{\star\mu \star\nu}=
g^{\mu\nu}\tilde{R}_{\star\mu ;\nu}$,
and recall Equation~(\ref{tildeR}) to further probe the
structure of the Maxwell current
\begin{equation}
J_\mu=e(a_\mu \tilde{R}
+\frac{1}{2}\tilde{R}_{;\mu}) ~.
\label{Mcurrent}
\end{equation}

An important question is then which conformal metrics
might admit a non-vanishing R.H.S. of Equation~(\ref{Mcurrent}),
or even better: Which geometries will not do so?
Apart from some special cases, e.g.,~conformal Schwarzschild
and Schwarzschild-deSitter metrics~\cite{C2}, the~general answer is still unknown.
We emphasize that the conservation of the co-vector
$J^\mu$ needs not be considered an~external constraint,
but rather be a legitimate consistency condition which
does not break local scale invariance.
This only requires though, as~noted earlier, the~replacement
of  $A_\mu$ by $A_\mu-\eta_{;\mu}$
\medskip

\item
By the same token, the~variation with respect to
$a_\mu$ leads to the field equation
\begin{equation}
f^{\mu\nu}_{~\star\nu}=j^\mu~.
\end{equation}

It takes some algebra though to establish the analogy
with the Maxwell current, and~verify that the Weyl
current is indeed proportional to $A_{\mu}$, and~is
given explicitly by
\begin{equation}
j_\mu=e (A_\mu \tilde{R}
+3A^\nu_{\star\nu\star\mu}) ~.
\end{equation}

\item
The co-divergence of the Maxwell vector field resembles
a dilaton, with~the formal definition being the coefficient
of $\tilde {R}$ in the Lagrangian Equation~(\ref{Rlinear}),
namely
\begin{equation}
\phi^2=\frac{1}{2}e A^\mu_{\star\mu}
=e(\frac{1}{2}A^\mu_{;\mu} -a_\mu A^\mu)~.
\label{phi}
\end{equation}

The fact that the roots of such a dilaton-like
configuration are electromagnetic in origin is a~natural consequence of the Weyl/Maxwell mutual
sourcing.
It is only in the intermediate stages, as~expressed by
the successive actions $I_{1,2}$, that $\phi$ becomes
an independent scalar dilaton field on its own merits.
Later on, as~demonstrated by actions $I_{3}$,
in analogy with the Higgs mechanism, Equation~(\ref{phi})
represents the vacuum of the~theory.

\item
Finally, imitating Holdom's $U(1)\otimes U^\prime(1)$
kinetic mixing, one may switch on the analogous scale/gauge
symmetric Weyl/Maxwell kinetic mixing~\cite{Davidson}
\begin{equation}
{\cal L}_\epsilon=\frac{1}{2}\epsilon
g^{\mu\lambda}g^{\nu\sigma}
F_{\mu\nu}f_{\lambda\sigma}~,
\label{epsilon}
\end{equation}
parametrized by some dimensionless coefficient
$\epsilon$.
No dramatic effects are expected as long as minimally
coupled charged scalar fields or fermion fields are not
introduced.
Once they do enter the theory,
reflecting the opposite transformation laws of
$A_\mu\rightarrow -A_\mu$ versus
$a_\mu\rightarrow +a_\mu$, the~discrete CP
symmetry gets explicitly violated.
\end{itemize}

To summarize, the~general idea of Weyl/Maxwell
mutual sourcing has been formulated on two
field theoretical levels.
They are: (1) A preliminary theory, free of fundamental
scalars and fermion fields, and~(2) A full Weyl-Dirac
variant theory  incorporating a genuine real dilaton
scalar field.
The~main message is that the Weyl/Maxwell mutual
sourcing term is necessarily spacetime curvature (not
just metric) dependent and inevitably Ricci linear,
thereby suggesting that a non-vanishing spacetime
curvature can in principle induce an electric current.
A central (and quite a novel) role is played in the
theory by the co-divergence of the Maxwell vector
field $A^\mu_{\star\mu}$.
In the basic version, prescribed by the action $I_0$
(see Equation~(\ref{Rlinear})),
serving as the coefficient of the Ricci curvature
term, it effectively resembles a dilaton field $\phi^2$
whose roots are thus counter intuitively electromagnetic
in origin.
The idea elegantly and most naturally fits into the
Weyl-Dirac (incorporating Brans-Dicke) theory.
Originally, the~latter exclusively allows for the quartic
potential term $\Lambda \phi^4$, but~in the
Weyl/Maxwell mutual sourcing extension, prescribed
by the action $I_{3}$, it is consistently
traded for the Higgs-like potential $\Lambda (\phi^{\dagger}
\phi-\frac{1}{2}e A^\mu_{\star\mu})^ 2$ without upsetting
the local scale invariance.
In other words, $A^\mu_{\star\mu}$ serves as (to be
referred to) a conformal vacuum expectation value.
Bearing in mind that a~spontaneous local scale symmetry
breaking mechanism is still very much at large, we can
only hope that the theory discussed may hopefully
contribute (currently under extensive investigation) in
this field theoretical~direction.

\acknowledgments{T.Y. was supported by the Israeli Science
Foundation Grant No. 1635/16.}

\end{document}